\documentclass{article}

\usepackage{arxiv}

\usepackage[utf8]{inputenc} 
\usepackage[T1]{fontenc}    
\usepackage{hyperref}       
\usepackage{url}            
\usepackage{booktabs}       
\usepackage{amsfonts}       
\usepackage{nicefrac}       
\usepackage{microtype}      
\usepackage{lipsum}

\usepackage{graphicx}
\usepackage{epstopdf}
\usepackage{amssymb}
\usepackage{amsmath}
\usepackage{multirow}
\usepackage{color}
\usepackage{tabularx,booktabs}
\usepackage{array}
\newcolumntype{P}[1]{>{\centering\arraybackslash}p{#1}}
\usepackage{booktabs, caption, makecell}

\usepackage{threeparttable}

\title{Connectivity-informed Drainage Network Generation using Deep Convolution Generative Adversarial Networks}

\author{
  Sung Eun ~Kim \thanks{saint.kse@gmail.com;  https://github.com/saint-kim/RiverDCGANs.}\\
  Water Resources Research Center \\
  Civil and Environmental Engineering\\
  University of Hawai'i at M\=anoa\\
  Honolulu, HI, 96822 USA \\
  \texttt{sekim7@hawaii.edu} \\
   \And
 Yongwon ~Seo \\
  Department of Civil Engineering\\
  Yeungnam University\\
  Gyeongsan, 38541 South Korea \\
  \texttt{yseo@ynu.ac.kr} \\
   \And
 Junshik ~Hwang \\
  Department of Civil Engineering\\
  Yeungnam University\\
  Gyeongsan, 38541 South Korea \\
  \texttt{jshwang@ynu.ac.kr} \\
   \And
 Hongkyu ~Yoon \\
  Geomechanics Department\\
  Sandia National Laboratories\\
  Albuquerque, NM, 87123 USA \\
  \texttt{hyoon@sandia.gov} \\
   \And
 Jonghyun ~Lee \\
  Water Resources Research Center \\
  Civil and Environmental Engineering\\
  University of Hawai'i at M\=anoa\\
  Honolulu, HI, 96822 USA \\
  \texttt{jonghyun.harry.lee@hawaii.edu} \\
}

\begin{document}
\maketitle

\begin{abstract}
Stochastic network modeling is often limited by high computational costs to generate a large number of networks enough for meaningful statistical evaluation.
In this study, Deep Convolutional Generative Adversarial Networks (DCGANs) were applied to quickly reproduce drainage networks from the already generated network samples without repetitive long modeling of the stochastic network model, Gibb's model.
In particular we developed a novel connectivity-informed method that converts the drainage network images to the directional information of flow on each node of the drainage network, and then transform it into multiple binary layers where the connectivity constraints between nodes in the drainage network are stored.
DCGANs trained with three different types of training samples were compared; 1) original drainage network images, 2) their corresponding directional information only, and 3) the connectivity-informed directional information.
Comparison of generated images demonstrated that the novel connectivity-informed method outperformed the other two methods by training DCGANs more effectively and better reproducing accurate drainage networks due to its compact representation of the network complexity and connectivity.
This work highlights that DCGANs can be applicable for high contrast images common in earth and material sciences where the network, fractures, and other high contrast features are important. 
\end{abstract}


\section{Introduction}

Runoff assessment has long been an important topic of hydrology for the purpose of water resources management, flood control, and ecological and environmental restoration. 
Runoff from a catchment primarily depends on two characteristics: the hydro-meteorological characteristics of rainfall and the watershed characteristics \cite{yen1969laboratory}. 
Among the watershed characteristics, drainage network topology is one of the most important factors that directly affects the hydrologic response of a watershed given the spatial and temporal rainfall distributions \cite{Singh_1998}. 
However, characterization of the drainage network topology is often hindered due to missing data and complex loops inside drainage network \cite{seo2012}. 
Furthermore, data acquisition and hydrologic analysis of actual drainage networks require time‐consuming processes. 

To overcome these difficulties in analyzing real drainage networks, statistical description of network topology has been utilized to generate drainage networks that can be used to assess the effect of drainage network topology on runoff \cite{seo2012,seo2013}.
Among many stochastic network generation models, Gibbs' model has been successfully used to perform hydrologic analysis of urban draining networks \cite{seo2012}. 
The Gibbs' model is a stochastic network generation model based on Gibbs' measure \cite{ising1925beitrag,kindermann1980markov} where the maximum entropy and a Markov random field \cite{kindermann1980markov} are used to define the complex network topology. 
For example, the Gibbs' model has been used to classify urban drainage network in Chicago areas \cite{seo2012,seo2013} and simulate alternative networks with similar hydrologic response \cite{seo2014a, seo2014b}.
However, it takes a relatively long time to generate a number of large networks enough for meaningful statistical evaluation because the Gibbs' model has to consider all possible flow directions at each node of the network\cite{barndorff_98}.
Hence, hydrologic analysis of networks generated with the Gibbs' model becomes less practical when quickly generating many large and complex networks.

Recent advances in deep learning methods \cite{lecun2015deep} can provide a promising approach to learning features underlying relationships (e.g., latent space) among data, classifying classes and labels, generating images and data, and scientific machine learning \cite{Baker_et_al_2019}. 
Among these techniques, generative adversarial networks (GANs) have a deep generative framework that can effectively learn a probability distribution of training sample data and generate realistic samples from the given distribution without explicitly modeling the probability density function \cite{goodfellow2014deep,Goodfellow2016NIPS2T}.
GANs have demonstrated remarkable results in image synthesis, image translation, data augmentation, and image/data reconstruction \cite{YI2019101552,9043519}.

Among various GANs implementations, two variants of GANs provide promising potential for generating the drainage network.
The first is deep convolutional GANs (DCGANs).
Radford et al. (2015)\cite{RadfordMC15} combines convolutional neural networks (CNNs) with GANs, \emph{i.e.}, DCGANs, to learn a hierarchical structure of image samples for better image representations.
Mosser et al. (2017)\cite{mosser17reconstruction} applied DCGANs with micro-computed tomography (micro CT) images to reconstruct the three-dimensional porous media and demonstrated the performance of the proposed DCGANs comparing with conventional geostatistical methods.
Recently, Kim et al. (2020)\cite{kimse2020} successfully applied DCGANs for generating the arbitrary large size of statistical realizations of two and three dimensional earth materials such as sphere packing and subsurface channels with various degree of connectivity and structural properties.
These studies showed that once trained, DCGANs can quickly generate/reconstruct multiple plausible images with various patterns that satisfy important statistical features of the training image samples with very low computational cost.
However, it does not guarantee that DCGANs can always reproduce the physical information (network complexity and connectivity) inherent in the original drainage network sample.
Moreover, the drainage network image has high-frequency features (i.e., a large contrast in the intensity of the neighboring pixel values in image data, such as points, lines, or graphs) which CNNs are often struggled to extract without carefully designed neural network architecture \cite{xu2019frequency}.

The second is conditional GANs (CGANs).
Mizra and Osindero(2014)\cite{MirzaO14cgan} introduced the conditional version of GANs (CGANs), which can be constructed by adding external information (tags or labels) to both training images and the generated images. 
This study showed that it is possible to control the properties of output image by conditioning the GANs model on additional information.
CGANs have been applied to various research for image synthesis with different conditional contexts such as categorical image generation, text-to-image synthesis, and semantic manipulation \cite{8411144, yang2018diversitysensitive, DBLP:journals/corr/abs-1711-11585}.
As demonstrated in many image processing and analysis examples \cite{YI2019101552}, DCGANs framework can be suitably combined with additional information like CGANs and generate samples with complex patterns and high-frequency features more reliably.

In this study, DCGANs were used as a deep learning framework to quickly generate many drainage networks with various patterns based on the drainage network samples already created by the stochastic network generation model, the Gibb's model.
Additionally, we proposed a novel connectivity-informed drainage network generation method to effectively train DCGANs with high frequency features and better reproduce accurate drainage networks.
The key idea of the proposed method is to convert the drainage network images to the directional information (right, left, up, down) of flow to the outlet from each node of the drainage network, and then transform the directional information into several binary layers where the contrast and connectivity of one node with neighboring nodes are stored effectively. In this way, the connectivity information of the network topology can be implicitly conditioned during the training of DCGANs. 
In the next section, GANs and DCGANs are briefly introduced, followed by the main concepts and advantages of the proposed connectivity-informed drainage network generation method are presented.
The training and network generation results are then compared to demonstrate the performance of the proposed method in the results and discussion section.
Finally, a summary of the important results of this study and possible future developments are described in the conclusion section.

\section{Methodology}

Here we briefly introduce Generative Adversarial Neural Networks (GANs) and Deep Convolutional GANs (DCGANs), followed by the connectivity-informed DCGANs developed in this study.

\subsection{Generative Adversarial Neural Networks (GANs)}

GANs introduced by Goodfellow et al. (2014)\cite{goodfellow2014deep} are one of deep neural networks with a new framework for estimating generative models via adversarial models. 
GANs train two models including a generative model $G$ and a discriminative model $D$.
The generative model captures the ``true'' data generation process for the training images, while the discriminative model determines whether samples are taken from either those generated by  $G$ or the training samples \cite{mosser17reconstruction}.
To approximate a generator distribution $p_{g}$ over ``true'' data x, the generator builds a mapping function $G(z;\theta_{g})$ in which a vector $z$ is generated from a prior noise distribution $p_{g}(z)$. $\theta_{g}$ represents parameters of $G$ and $z$ is typically a Gaussian random vector.
The discriminator, $D(x;\theta_{d})$, yields a single scalar ($D:\mathbb{R}^{n}\rightarrow [0, 1]$) representing the probability of the data $x$ originating from training samples rather than those from $G(z;\theta_{g})$ \cite{goodfellow2014deep,Goodfellow2016NIPS2T}. 
Then, two models contest with each other in a game framework such that the $G$ model learns ``true'' data generating process to deceive the $D$ model while the $D$ model distinguishes the true data from the $G$ model-generated samples as in following optimization problems:

\begin{equation}
    \min J^{(D)} = -\frac{1}{2}\Big\{\mathrm{E}_{x \sim P_{true}(x)}[log D(x)]+\mathrm{E}_{z \sim P_{g}(z)}[log(1- D(G(z)))]\Big\}  
    \label{eq:JD}
\end{equation}
\newline  
\begin{equation}
    \min J^{(G)} = -\frac{1}{2}\mathrm{E}_{z \sim P_{g}(z)}[log(D(G(z)))] 
    \label{eq:JG2}
\end{equation}

\begin{figure}[hb]
    \centering
    \includegraphics[width=0.9 \linewidth]{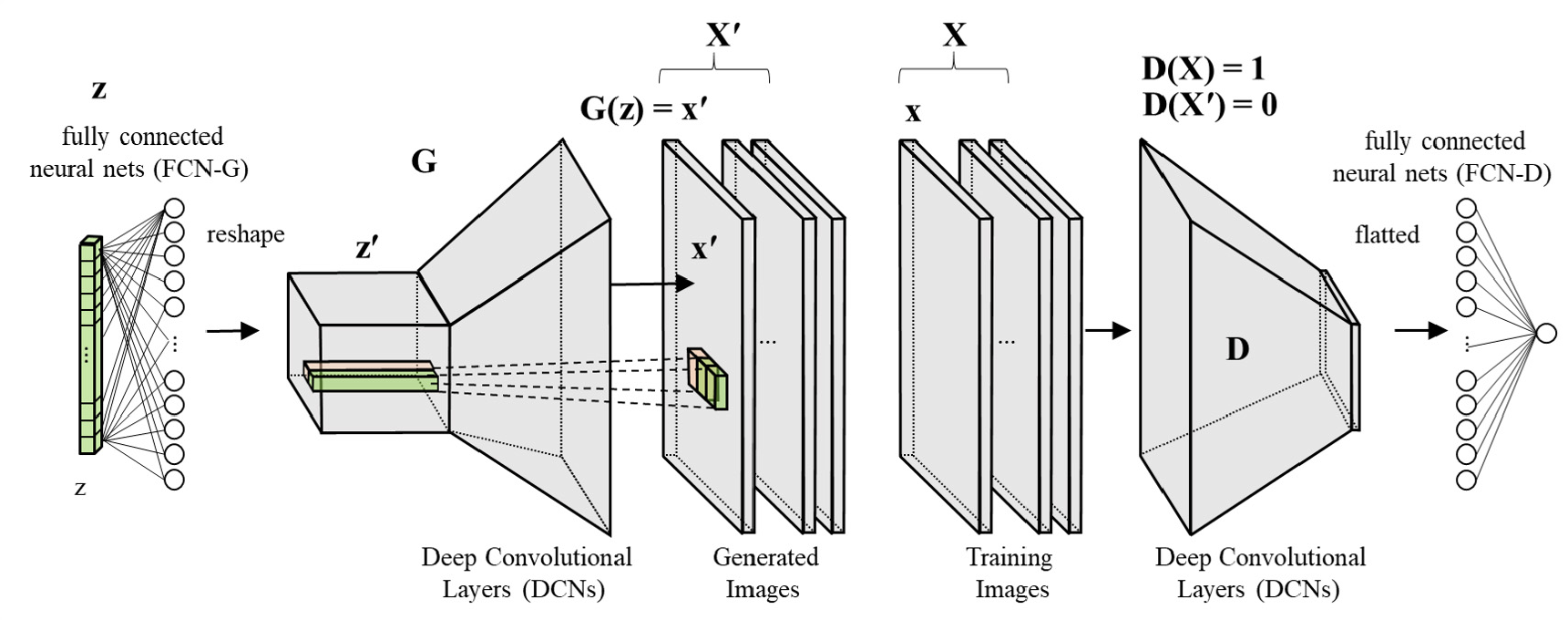}
    \caption{Schematic representation of DCGANs: the deep convolutional neural networks were utilized to develop the GANs.}
    \label{fig:DCGANs_scheme}
\end{figure}

The models $G$ and $D$ are trained simultaneously. 
Parameters $\theta_{d}$ of the $D$ model are adjusted to minimize $J^{(D)}$ for the $D$ model to distinguish between the real and the $G$ model generated (Eq.~\ref{eq:JD}) while parameters $\theta_{g}$ of the $G$ model are adjusted to minimize $J^{(G)}$ for the discriminator being correct (Eq.\ref{eq:JG2}). 
This results in $G$ model trained to make the value of $D(G(z))$ close to 1 and $D$ model trained to make the value of $D(G(z))$ to $0$. 
Through this dual optimization procedure, GANs can approximate the generator asymptotically to the true data generating process confirmed by the discriminator as shown in Fig.\ref{fig:DCGANs_scheme}. 

\subsubsection{Deep Convolutional GANs (DCGANs)}

GANs are often unstable in training, resulting in the generator that produces nonsensical outputs and/or mode collapsing, \emph{i.e.}, a limited diversity in generated samples \cite{Goodfellow2016NIPS2T}.
Mode collapse is an inherent problem in the training procedure of GANs \cite{srivastava2017veegan}.
The most effective way to reduce the mode collapse problem is to use a better classifier for training all modes of data distribution \cite{srivastava2017veegan, bang2018mggan}.
Convolutional neural networks (CNNs) are regarded as a better option for classification than fully-connected neural networks and able to identify useful representations and features of the inputs. 
CNNs can also contain more complex features into the neural network architecture.
Therefore, the DCGANs have been developed to utilize the deep convolutional neural networks in the GANs since the representation of the learned data distribution can be stored in convolutional layers efficiently, which is reused to generate samples.
This convolutional nature in CNN enables GANs to generate many samples similar to the training sample with computational efficiency.

\subsubsection{Architecture of DCGANs}

It is widely recognized that   the identification of the appropriate DCGANs architecture for the optimal training would require extensive model exploration. 
Radford et al. (2015)\cite{RadfordMC15} identifies a family of deep convolution architectures that results in stable training across a range of samples, which allows us to train higher resolution and deeper generative models. 
In this work, we adopt main architectural features from Radford et al. (2015)\cite{RadfordMC15}: 1) stride convolutions instead of any pooling layers , 2) batch normalization in both the generator and discriminator, 3) no hidden layers in fully connected net in both the generator and discriminator, 4) ReLU activation in the generator for all layers except for the output, which uses the Tanh activation funcation, and 5) LeakyReLU activation in the discriminator for all layers except for the output, which uses the sigmoid activation function.
\begin{figure}[hb]
    \centering
    \includegraphics[width=0.9\linewidth]{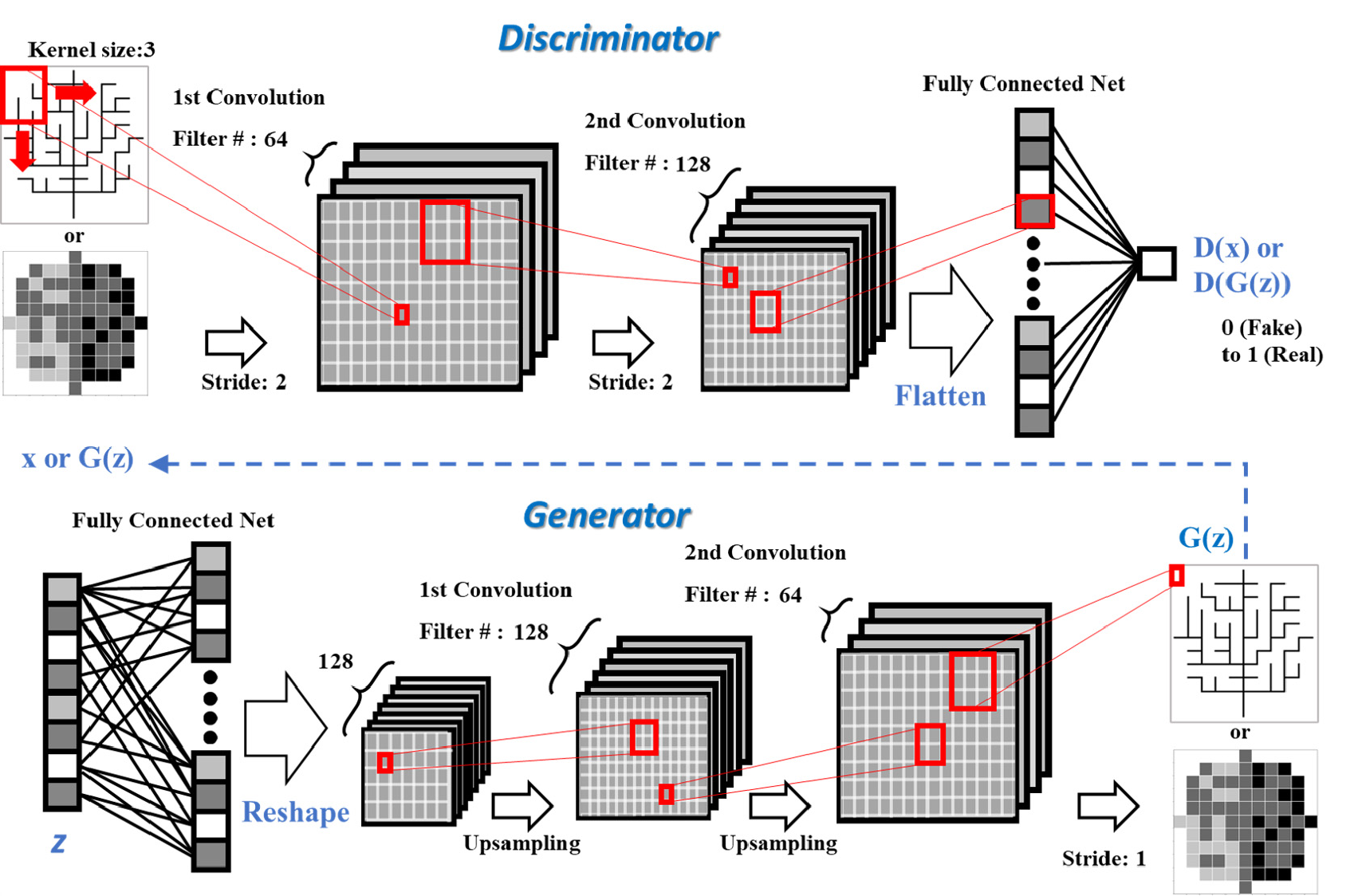}
    \setlength{\belowcaptionskip}{-8pt}
    \caption{Architecture of DCGANs used in this study. The generator and discriminator are designed to have a symmetrical structure composed of fully connected neural nets and two convolution layers with a kernel size of (3, 3).}
    \label{fig:DCGANs_Archi}
\end{figure}

The architecture of DCGANs used in this study is shown in Fig.~\ref{fig:DCGANs_Archi}.
The $D$ model is composed of two forward stride/two convolution layers with a kernel size of three, and the output was converted as a probability (False as 0 to True as 1) using the sigmoid activation function.
In the $G$ model, the latent vector $z$ with a dimension of 100 was drawn from a Gaussian distribution.
A fully connected neural network (FCN) was reshaped into a 4-dimensional tensor that was used as the start of the convolution stack.
Two backward upsampling convolution layers with a kernel size (3, 3) were used, and the output was converted by a forward stride convolution layer with kernel size (3, 3) into the same size image as the training image.
Both $D$ and $G$ models were trained by the adaptive momentum estimation (Adam) optimization algorithm with a starting learning rate of 0.0002 with a momentum ($\beta_{1}$) of 0.5, and a total of 100,000 epochs with a mini-batch of size 64.
A dropout with the probability of 0.25 was applied to both the generator and discriminator.
A small value of both learning rate and momentum was used for the stabilized training and the convergence of the model.
A larger number of epochs has been applied to analyze the change of the loss value in the training of the generator and discriminator.
The loss of GANs was estimated by the binary cross-entropy function.
The parameter values used in this study are presented in Table.~\ref{table:para}. 
\begin{table}[t]
\centering
\caption{Parameter Values of DCGANs used in this study}
\renewcommand{\arraystretch}{0.9}
\scalebox{0.9}{
\begin{tabular}{cll}
\hline
\multicolumn{1}{l}{\textbf{}} & \multicolumn{2}{c}{\textbf{ Parameters \& Values}} \\ \hline
\textbf{\begin{tabular}[c]{@{}c@{}}Latent space\\ (z dimension)\end{tabular}} & \multicolumn{2}{c}{100} \\ \hline
\multirow{2}{*}{\textbf{Convolution layer}} & \textbf{Generator} & 128 / 64 filters with kernel size = 3 \\
 & \textbf{Discriminator} & 64 / 128 filters with kernel size = 3 \\ \hline
\multirow{4}{*}{\textbf{Optimizer}} & \multicolumn{2}{c}{\textbf{Adam with mini-batch}} \\ \cline{2-3} 
 & \textbf{Learning rate} & 0.0002 \\
 & \textbf{Momentum} & $\beta _{1} = 0.5$, \hspace{0.5cm} $\beta _{2} = 0.999$ \\
 & \textbf{Batch size} & 64 \\ \hline
\multirow{2}{*}{\textbf{Regularization}} & \textbf{Generator} & Batch normalization with a momentum of 0.8 \\ \cline{2-3} 
 & \textbf{Discriminator} & \begin{tabular}[c]{@{}l@{}}Dropout with 25 \%, \\ Batch normalization with a momentum of 0.8\end{tabular} \\ \hline
\multirow{2}{*}{\textbf{Activation function}} & \textbf{Generator} & \begin{tabular}[c]{@{}l@{}}ReLu,\\ Tanh (output layer)\end{tabular} \\ \cline{2-3} 
 & \textbf{Discriminator} & \begin{tabular}[c]{@{}l@{}}LeakyReLu (alpha = 0.2),\\ Sigmoid (output layer)\end{tabular} \\ \hline
\textbf{Loss function} & \multicolumn{2}{l}{Binary Cross-entropy} \\ \hline
\end{tabular}
}
\label{table:para}
\end{table}

\subsection{Generation of Training Images}

We applied the drainage network images with two different network complexities to evaluate how accurately DCGANs with and without the connectivity-constrained information capture and reproduce the complexity and connectivity in training samples.
The drainage network training images with a specific network complexity were generated using the Gibbs' model.
To generate a dendritic network with the Gibbs' model, a Markov chain is defined with the spanning trees of $S$ as the state space. Let a tree, $s$ belong to a set of trees, $S$ and two trees $s_{1}$ and $s_{2}$ be adjacent. 
The transition probability from $s_{1}$ to $s_{2}$ is defined as follows \cite{troutman1992gibbs}:

\begin{equation}
R_{s_1s_2} =\begin{cases} r^{-1} \min{ \left[ 1,\exp\left(-\beta \left( H(s_2) - H(s_1) \right)\right)\right] }, & s_2 \in N(s_1) \\
1-\sum_{s\in N(s_1)} R_{s_1 s_2}, & s_2 = s_1 \\ 0, & \mbox{otherwise} \end{cases}
\end{equation}

where $N(s_1)$ is the set of trees adjacent to $s_1$, and $\beta$ is a parameter that represents the extent to which the sinuosity of the network is reflected in the generation of the new spanning tree, $s_2$. 
\begin{figure}[hb]
    \centering
    \includegraphics[width=1.0\linewidth]{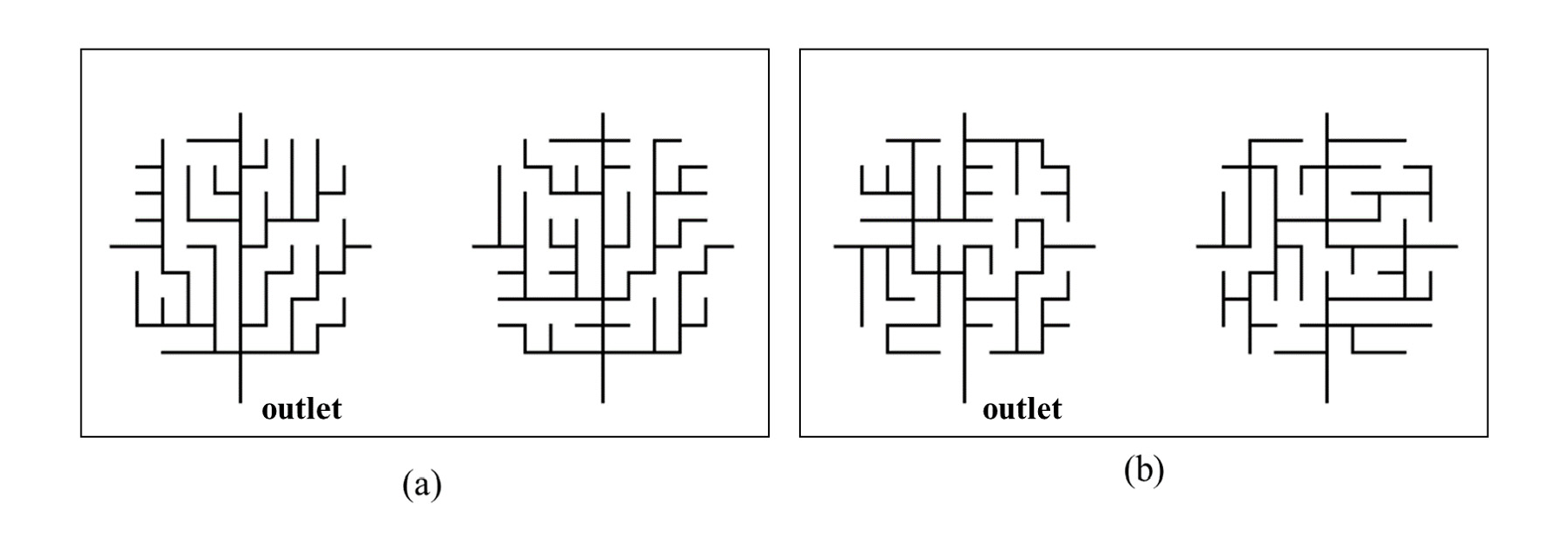}
    \caption{Drainage Networks generated by Gibbs’ model with (a) $\beta = 10^{3}$ to generate simple network topology to the drainage outlet (down point) and (b) $\beta = 10^{-4}$ to generate complex network topology to the drainage outlet} 
    \label{fig:network_model}
\end{figure}
 
Depending on the value of a parameter $\beta$, the Gibbs’ model can generate the drainage network images with specific network complexity.
For example, when $\beta$ is equal to zero, the overall sinuosity of a network has no relationship to the transition probability and the transition probabilities are the same in all possible directions, which produces a high sinuosity in the generated networks.
In this study, we trained DCGANs with drainage network images generated using the Gibbs’ model with two different $\beta$ values ($10^{3}$ and $10^{-4}$) to represent simple and complex drainage network topologies (Fig.~\ref{fig:network_model}).
The case of  $\beta$ = $10^{3}$ allows the network to generate in the three different directions (left, right, and downward), while the case of $\beta = 10^{-4}$ generates the network in the four different directions (left, right, upward, and downward)  resulting in more complex pathway to the drainage outlet.

\subsection{Connectivity-informed Training Images}

The drainage network is a fully connected network with high-frequency features consisting of points and lines (Fig.~\ref{fig:network_model}).
CNNs are generally good at extracting the ``texture'' information from an imagery data, but they are often not good at estimating complex and sparse features such as points, lines, and graphs without careful architecture designs \cite{xu2019frequency}. Additional information combined with an effective neural network architecture that can suitably extract key image patterns should be implemented to improve the training efficiency of a CNN-based generation model.
Hence, we use explicit transformations of the original data format achieving dimension reduction and enhancing the directional connectivity information to effectively learn complex high-frequency features of the drainage network images and reproduce the network complexity and connectivity. To assess the effectiveness of the proposed approach during the training, we will test two different approaches using one with only directional information matrix and the other with both directional and connectivity information.

\subsubsection{Conversion of Network Image to Directional Information}

\begin{figure}[hb]
    \centering
    \includegraphics[width=0.6\linewidth]{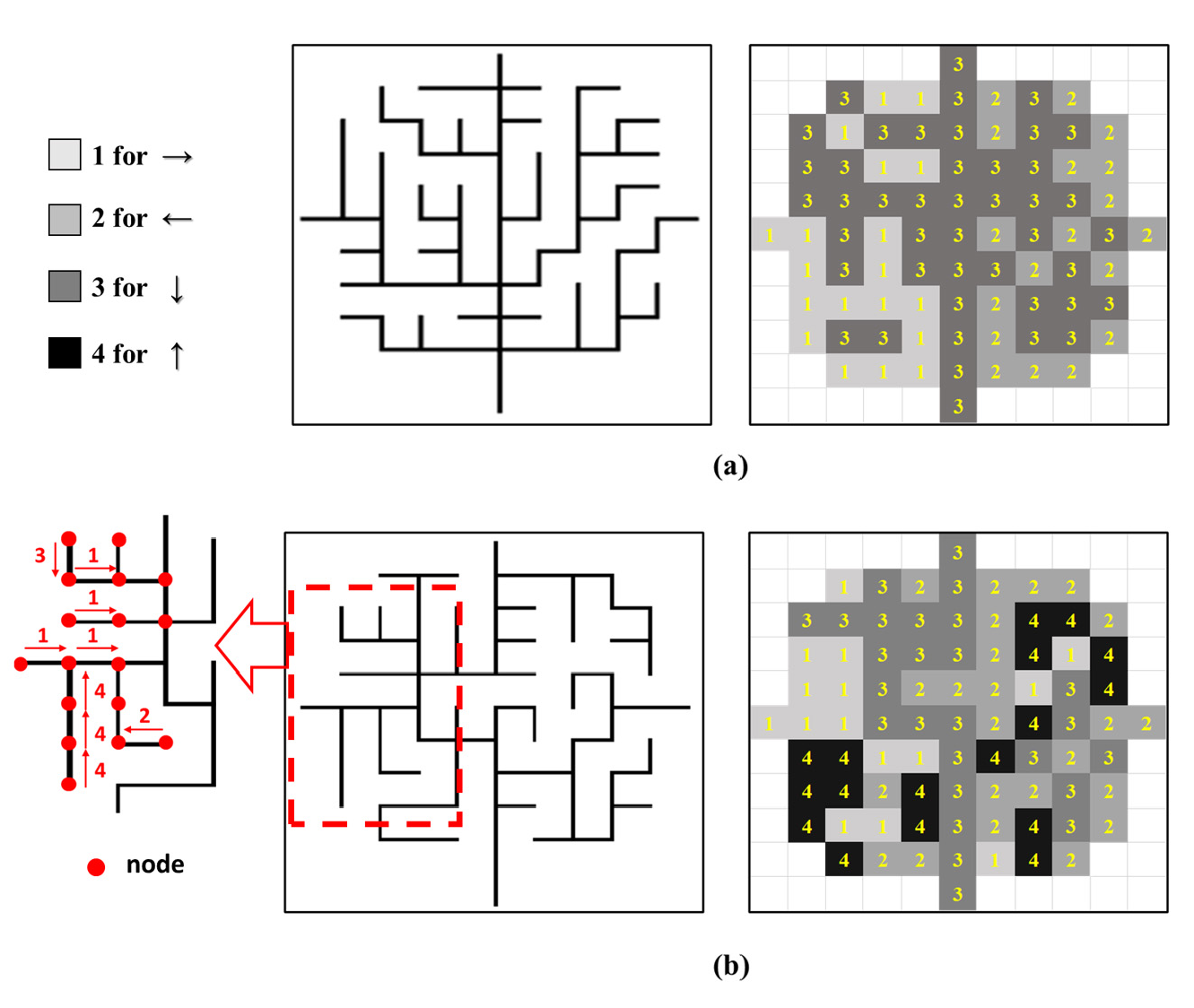}
    \caption{Drainage network images and the corresponding $\mathfrak{D}$-matrix for (a) $\beta = 10^{3}$ and  (b) $\beta = 10^{-4}$. The drainage network images (size of 120 $\times$ 120 pixels) were converted into the $\mathfrak{D}$-matrices (size of 11 $\times$ 11) which have the directional information only at the node.}
    \label{fig:inputs}
\end{figure}

We propose to convert the directional information of the drainage network into ``$\mathfrak{D}$-matrix'' that compactly represents the direction of flow to the outlet of the drainage network as shown in Fig.~\ref{fig:inputs}.
The $\mathfrak{D}$-matrix extracts the directional information of the drainage network along each drainage segment where the index `1' indicates the direction of flow to the right `$\rightarrow$' at the node, `2' for the left flow `$\leftarrow$' ,`3' for the downward flow `$\downarrow$', and `4' for the upward flow `$\uparrow$'.
It is also straightforward to transform the $\mathfrak{D}$-matrix back to the network image inversely.
Another advantage of using the $\mathfrak{D}$-matrix is the dimension reduction to (11 $\times$ 11) from a size of the drainage network image (120 $\times$ 120) since the $\mathfrak{D}$-matrix contains the directional information only along the drainage path. 
This can reduce the computational cost for training in DCGANs significantly.
However, this $\mathfrak{D}$-matrix may not allow DCGANS to learn the crucial spatial structure in the fully connected drainage networks.
Without accounting for the connectivity between adjacent nodes, each node in the $\mathfrak{D}$-matrix allows any directions regardless of the direction of the adjacent nodes, which may lead to a low prediction performance of DCGANs using only the $\mathfrak{D}$-matrix.

\subsubsection{Connectivity-informed Directional Information}

\begin{figure}
    \centering
    \includegraphics[width=0.9\linewidth]{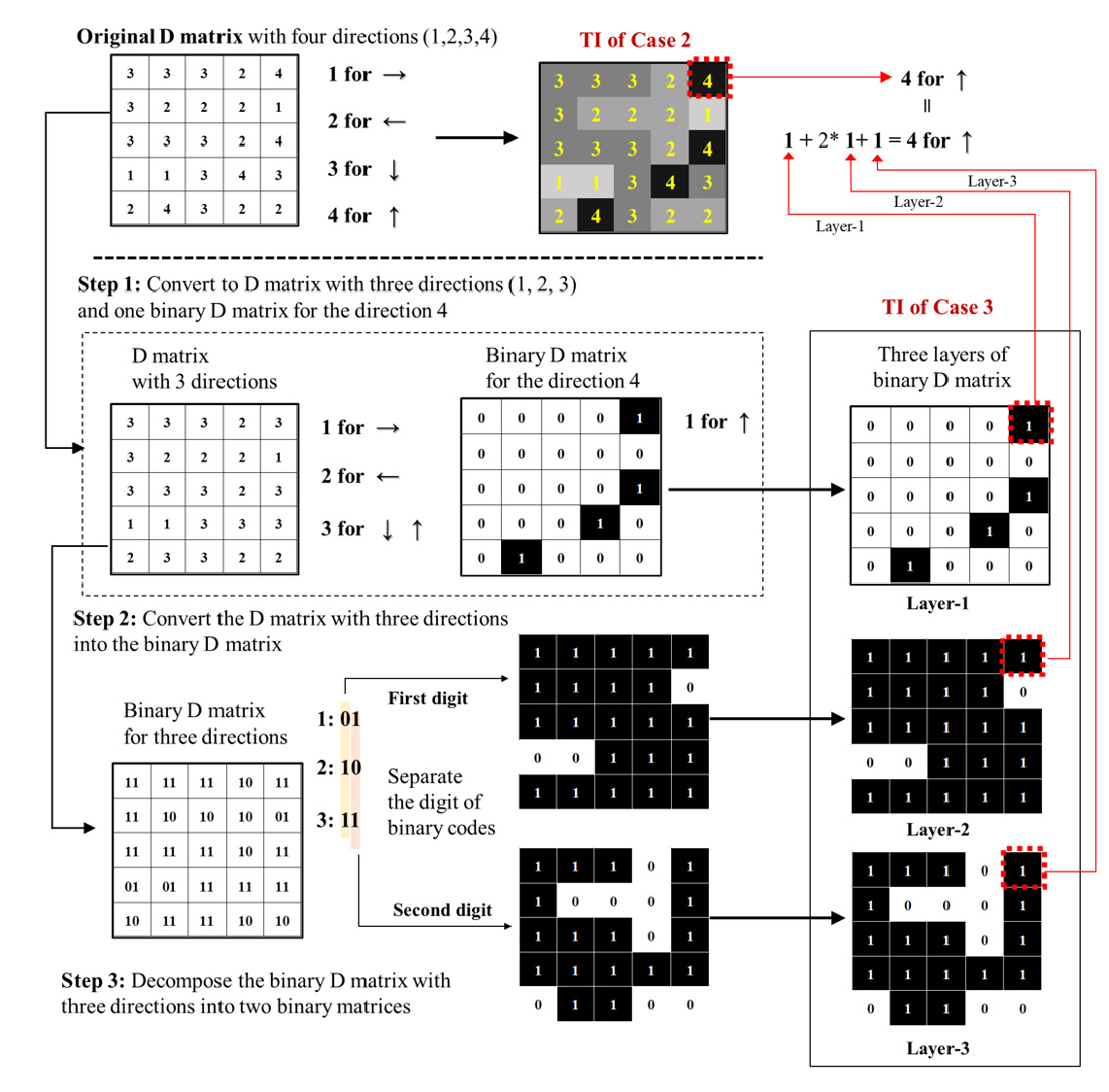}
    \caption{Conversion of the $\mathfrak{D}$-matrix into the connectivity-informed $\mathfrak{D}$-matrix. Constraints of the directional information on connectivity were achieved by separating the $\mathfrak{D}$-matrix into several layers of binary matrices for the directional information to provide physics-informed constraints between different directions.}
    \label{fig:input_D}
\end{figure}

Next, local direction information and associated network connectivity are incorporated for connectivity-informed learning.
In particular, constraints on the node direction ensuring the network connectivity were implicitly imposed by separating the $\mathfrak{D}$-matrix into two or three binary matrices (or channels) for each direction so that spatial patterns in each direction are better trained as shown in Fig.~\ref{fig:input_D}. 
Specifically, upward direction at each node is first stored in a binary channel matrix (Layer-1) to indicate whether there is upward direction at the node as 1 or not as 0. 
Then left and right directions are stored in the second and third binary channel matrices (Layer-2 and Layer-3), respectively in a similar way. 
If left and right directions coexist in both second and third matrices, it represents the downward direction.
By doing so, convolutional filters in DCGANs will learn the patterns of left and right dominant flow directions and their spatial connectivity from each channel matrix and if those two directions coexist in the two matrices, downward gravity direction will be selected.
As an example, in a less complex drainage network with dominantly three directional flows (indices = 1,2,3) without or with a very low probability of the upward directional flow (index = 4) as shown in Fig.~\ref{fig:network_model} (a), the direction and connectivity information can be decomposed with its binary numbers into two new binary matrices with the right-flow dominant area occupied by the indices `1' and `3' (right and downward flow directions) as 1 in the first channel matrix and the left-flow dominant area occupied by the indices `2' and `3' (left and downward flow directions) as 1 in the second channel matrix (see Layer-2 and Layer-3 in Fig.~\ref{fig:input_D}). 
The two new binary matrices then perform an element-wise logical operation to determine the left (`10'), right (`01') or downward (`11') direction. 
These two different but overlapping areas explain the key features of spatial network patterns which would work as a soft physics constraint to reproduce the fully connected drainage network better. 
This is the most important aspect of the connectivity-informed drainage network method proposed in this study; the directional connectivity information is stored as the decomposed binary matrices so that the spatial connectivity information can be properly extracted through the deep convolutional networks.
Note that for the simple drainage network topology, only two layers (Layer-2 and Layer-3 in Fig.~\ref{fig:input_D}) for the left, right and downward directions will be needed. 
Complex drainage network with all four directional flows as shown in Fig.~\ref{fig:network_model} (b) will require one more layer (Layer-1) for the additional downward direction. 
More detailed network with 8 or 16 directions can use the propose transformation suitably with increasing number of channel matrices.
\subsubsection{Experimental Cases}

To demonstrate the performance of DCGANs with the proposed methods, three cases with different training samples were created in this study.
In Case 1, the drainage network from the Gibbs' model with a size of 120 $\times$ 120 images was used for training DCGANs.
The generator created the same size of the drainage network images which were fed into the discriminator as described previously in the Methodology section.
In Case 2, the $\mathfrak{D}$-matrix data corresponding to the drainage network images used in Case 1 was used for training DCGANs.
Note that DCGANs were trained to generate the $\mathfrak{D}$-matrix of a size of 11 $\times$ 11 image. 
In Case 3, the connectivity-informed $\mathfrak{D}$-matrices were used as the training samples.
$\mathfrak{D}$-matrices with 2 (or 3) binary layers of a size of 11 $\times$ 11 $\times$ 2 (or 3) channel images were generated in this case.
The generated $\mathfrak{D}$-matrices in Cases 2 and 3 were transformed back to the corresponding drainage network and then compared with the drainage networks generated in Case 1. 
Subcases 1 and 2 of each Case (e.g., Cases i-1 and i-2 where i=1,2,3) represent the drainage network samples from the Gibbs' model with $\beta = 10^{3}$ (Case i-1) and $\beta = 10^{-4}$ (Case i-2), respectively.
Details of training samples used in each case are provided in Table~\ref{table:inputs}.

\begin{table}[h]
\centering
\caption{Training samples used in each case: (Case 1) the drainage network images from Gibbs' model, (Case 2) their corresponding $\mathfrak{D}$-matrix, and (Case 3) connectivity-informed $\mathfrak{D}$-matrices}
\scalebox{1.2}{
\begin{small}
\begin{tabular}{cclll}
\hline
\textbf{} & \textbf{\textbf{Subcase}} & \multicolumn{1}{c}{\textbf{Gibbs' model}} & \multicolumn{1}{c}{\textbf{Type of training samples}} & \multicolumn{1}{c}{\textbf{Size}} \\ \hline
\multirow{2}{*}{\textbf{Case 1}} & \textbf{1} & $\beta$ = $10^{3}$ & \multirow{2}{*}{Drainage Image} & \multirow{2}{*}{$120 \times 120$} \\
 & \textbf{2} & $\beta$ = $10^{-4}$ &  &  \\ \hline
\multirow{2}{*}{\textbf{Case 2}} & \textbf{1} & $\beta$ = $10^{3}$ & \multirow{2}{*}{$\mathfrak{D}$-matrix} & \multirow{2}{*}{$11 \times 11$} \\
 & \textbf{2} & $\beta$ = $10^{-4}$ &  &  \\ \hline
\multirow{2}{*}{\textbf{Case 3}} & \textbf{1} & $\beta$ = $10^{3}$ & $\mathfrak{D}$-matrix with 2 layers & $11 \times 11 \times 2$ \\
 & \textbf{2} & $\beta$ = $10^{-4}$ & $\mathfrak{D}$-matrix with 3 layers & $11 \times 11 \times 3$ \\ \hline
\end{tabular}%
\end{small}}
\label{table:inputs}
\end{table}

\section{Results and Discussion}

\subsection{Training Results}

\begin{figure}[hb]
    \centering
    \includegraphics[width=0.8\linewidth]{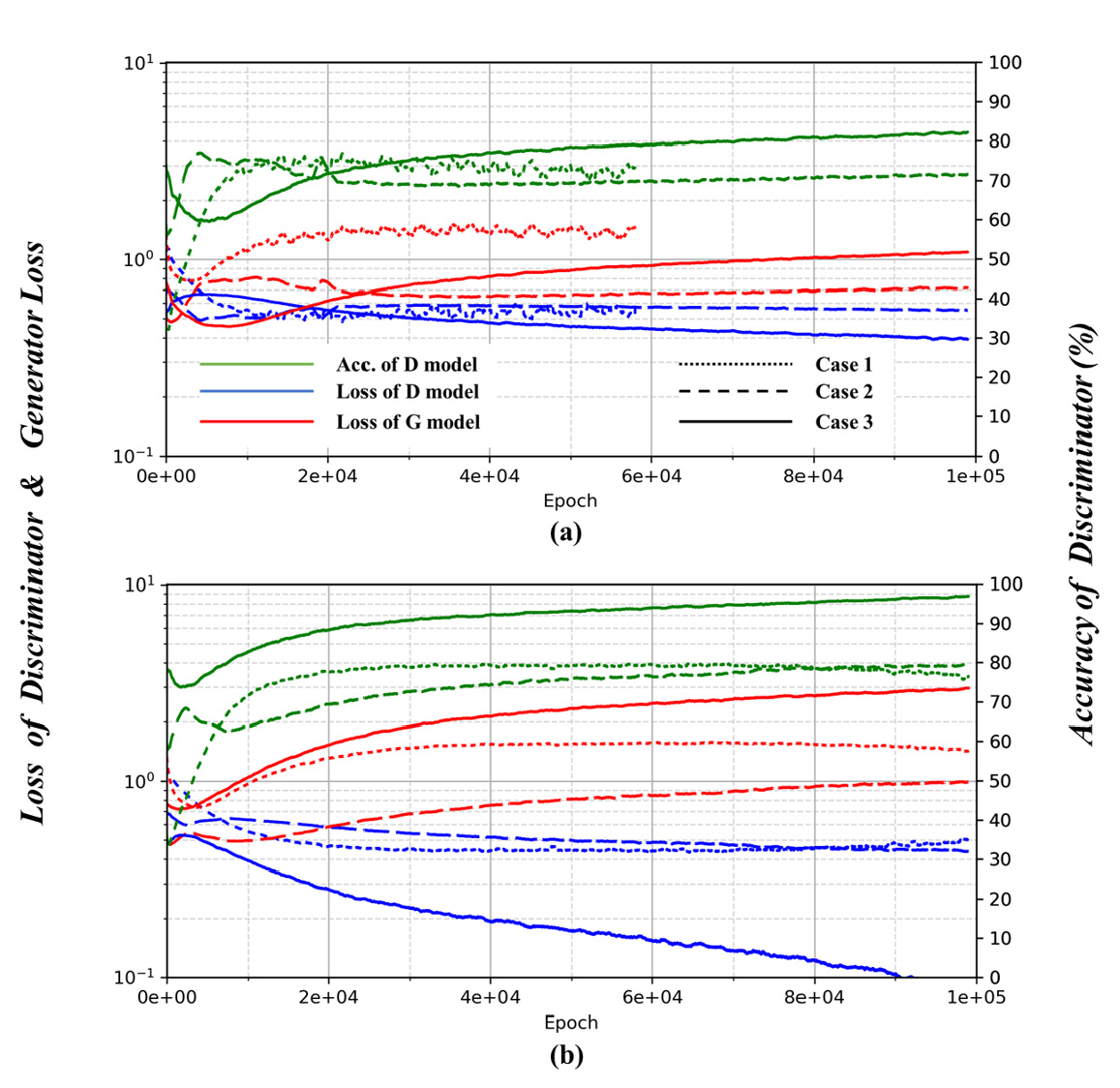}
    \caption{The change of loss values of the generator ($G$ model) and the discriminator ($D$ model) and accuracy of the discriminator over epochs for (a) Subcase 1 (Case 1-1, Case 2-1, Case 3-1) (b) Subcase 2 (Case 1-2, 2-2, 3-2). Note that Subcase 1 is with the less complex network and Subcase 2 with the more complex network (see Fig.~\ref{fig:network_model})}
    \label{fig:tr_result}
\end{figure}

In this study, DCGANs were trained with three different types of training samples, \emph{i.e.}, 120 $\times$ 120 images (Case 1), 11 $\times$ 11 $\mathfrak{D}$-matrix (Case 2), and 11 $\times$ 11 with 2 or 3 layers of $\mathfrak{D}$-matrices (Case 3) as in Table.~\ref{table:inputs}.
The shape (size) of the training sample affects the architecture and training of DCGANs.
Depending on the size of the training sample, the size of the first reshaped tensor right after the Fully Connected Net in the Generator in the Fig.~\ref{fig:DCGANs_scheme} or the number of layers and filters of CNNs should be adjusted in DCGANs.
To compare the training results for the different training samples (Table~\ref{table:para}) consistently, we set the size of the reshaped tensor to be scaled without changing other architectural parameters of DCGANs.
For the comparison of the training results, the loss values of the generator $G$ ($J^{G}$; red line) and the discriminator ($J^{D}$; blue line) with the accuracy of the $D$ model (green line) are compared in Fig.~\ref{fig:tr_result}.
The accuracy of the $D$ model represents the probability of the $D$ models for recognizing the real sample (\emph{i.e.}, training sample) and rejecting the generated sample correctly.
The moving average over 1,000 epochs were calculated to smooth the oscillating patterns of the loss values.

After some initial epochs in all loss graphs of Fig.~\ref{fig:tr_result}, it is shown that the loss function $J^{D}$ (blue line) decreases while $J^{G}$ (red line) increases in order to seek an equilibrium between the model $G$ and $D$.
In GANs, both $G$ and $D$ models were trained to minimize their loss functions in Eqs.~\ref{eq:JD} and~\ref{eq:JG2} and due to their adversarial relationship, the $G$ model learns indirectly only through the interaction with the $D$ model for the training samples,  leading to a typical convergence behavior as in Fig.~\ref{fig:tr_result}. 
It is worth noting that in initial epochs of the training, the $D$ model is not trained enough for distinguishing the real sample from the generated sample by the $G$ model; $J^{D}$ is higher than $J^{G}$, and the accuracy (green line) is approximately between 30\% and 70\%.
After initial epochs of the immature training, $J^{D}$ tends to decrease and the accuracy increases, \emph{i.e.}, the $D$ model gets smarter and the $G$ model is trained gradually following the updated $D$ model with a deceiving ratio.
This fine-tuning process makes the $G$ model more sophisticated and robust to reproduce plausible drainage networks.
Overall, the $D$ model minimizes the loss value more than the $G$ model and successfully rejects generated samples with high confidence as the number of epochs increases.

In Case 1 with the drainage network images as the training sample, the initial number of epochs for the immature training is small and the $D$ model does not get improved (i.e., the loss value does not decrease) with increasing the number of epochs.
In particular, in Case 1-1 (dotted line in Fig.~\ref{fig:tr_result} (a)), DCGANs' training was stopped due to unstable training at ~ 60,000 epochs due to the gradient vanishing problem.
In Case 2 using the $\mathfrak{D}$-matrix (dashed line in Fig.~\ref{fig:tr_result}) alone, the initial number of epochs corresponding to the immature training is less than Case 1, while the loss value of the $D$ model slowly decreases with increasing the accuracy ($\sim$70\% and 80\% for Cases 2-1 and 2-2, respectively) as epochs increase.
Case 3 (solid line in Fig.~\ref{fig:tr_result}) using the proposed connectivity-informed $\mathfrak{D}$-matrices has relatively longer initial epochs of the immature training and the loss value of the $D$ model decreases lower than other cases with higher accuracy ($\sim$80\% and 95\% for Cases 3-1 and 3-2, respectively) as epochs increase.
Those results indicate that with the proposed connectivity-informed approach, the $G$ models can be better trained through the improved $D$ models and longer immature training may result from the process of fine-tuning parameters to extract important network properties. 

\subsection{Drainage Network Generation}
\subsubsection{Network Connectivity}
\begin{figure}
    \centering
    \includegraphics[width=1.0\linewidth]{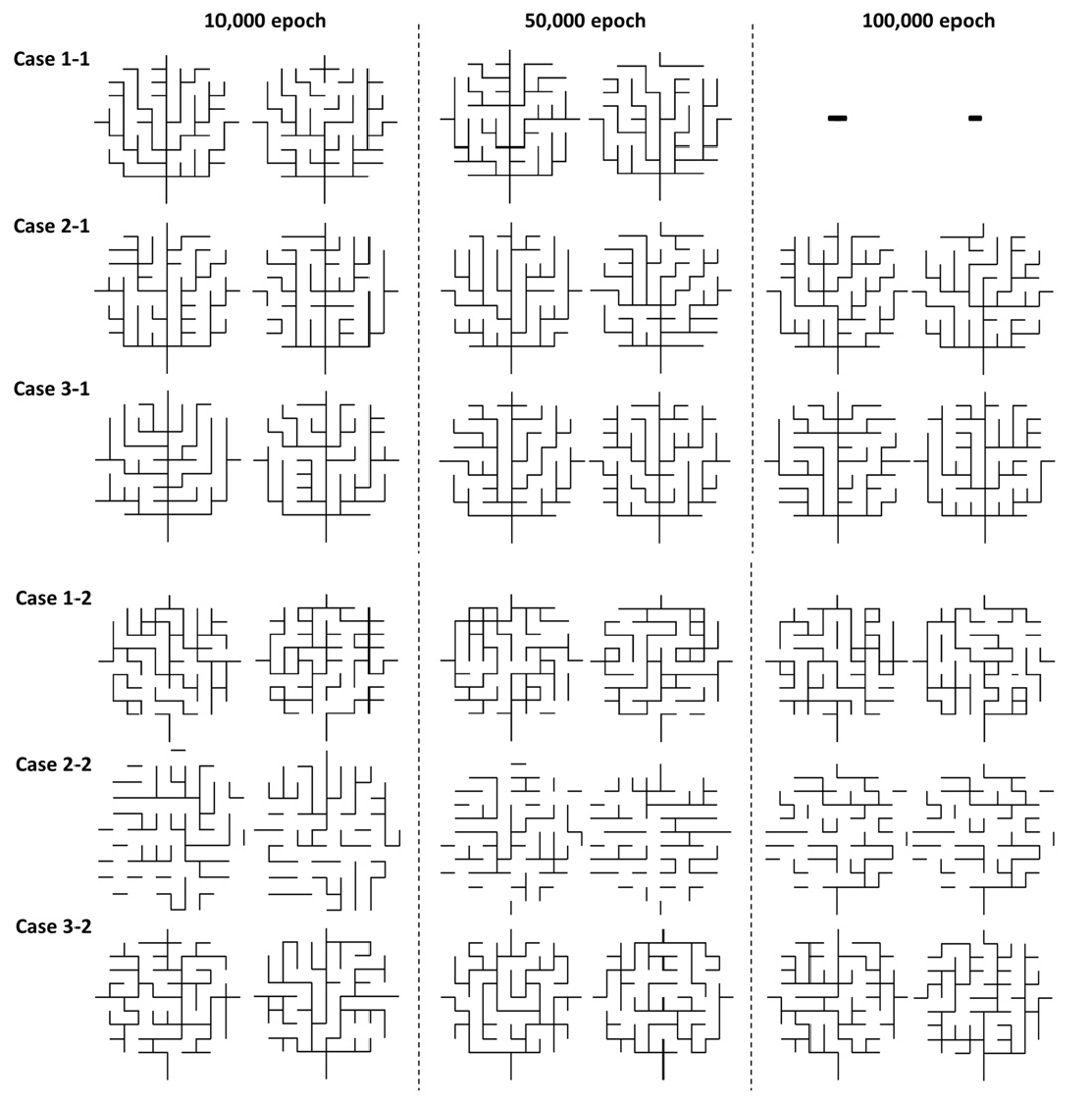}
    \caption{Comparison of generated drainage networks for six cases at 10,000, 50,000 and 100,000 epochs. Note that Subcase 1 (Case 1-1, 2-1, 3-1) is with the less complex network and Subcase 2 (Case 1-2, 2-2, 3-2) with the more complex network. Cases 1-2 and 2-2 rarely generated the fully connected drainage networks although the training results (\ref{fig:tr_result}) showed that both $D$ and $G$ models were trained and the accuracy of the $D$ model increases as epochs increase.}
    \label{fig:Gen}
\end{figure}

Since the drainage networks are fully connected networks, it is important not only to generate the similar structure of the drainage network in a shape, but also to reproduce the connectivity of the drainage network.
By fully connected network, we mean here that the network is acyclic and every node except outlet nodes have at least one downstream node.
To evaluate the performance on reproducing the full connectivity of the drainage network in each case, the number of fully connected drainage networks among the generated 10,000 drainage network samples were measured in Table~\ref{table:per}. A subset of generated drainage network samples are presented in Fig.~\ref{fig:Gen}.
\begin{table}[h]
\centering
\caption{Drainage network generation performance on connectivity reproduction and computational training efficiency. The number of fully connected drainage networks among the generated 10,000 drainage network samples were measured at four different numbers of epochs.}
\begin{threeparttable}
\scalebox{1.0}{
\begin{tabular}{lP{1.25cm}P{1.25cm}P{1.25cm}P{1.25cm}cc}
\hline
\multirow{2}{*}{} & \multicolumn{4}{c}{\textbf{\begin{tabular}[c]{@{}c@{}} \#\ of the fully connected drainage \\ networks at the epoch of \end{tabular}}} & \multicolumn{2}{c}{\textbf{\begin{tabular}[c]{@{}c@{}}Averaged\\ Time (Sec.)\end{tabular}}} \\ \cline{2-7} 
 & \textbf{1e4} & \textbf{2e4} & \textbf{5e4} & \textbf{10e4} & \textbf{\begin{tabular}[c]{@{}c@{}}Training\\ per 10 epoch\tnote{1}\end{tabular}} & \textbf{\begin{tabular}[c]{@{}c@{}}Generating\\ 10,000 samples\tnote{2}\end{tabular}} \\ \hline
\textbf{Case 1-1} & 3,566 & 4,163 & 5,697 & - & 7.47 & 43.34 \\ \hline
\textbf{Case 2-1} & 1,954 & 4,865 & 8,885 & 8,690 & 1.35 & 21.12 \\ \hline
\textbf{Case 3-1} & 9,262 & 9,216 & 9,134 & 9,123 & 1.25 & 15.03 \\ \hline
\textbf{Case 1-2} & 50 & 49 & 68 & 54 & 7.67 & 43.43 \\ \hline
\textbf{Case 2-2} & 0 & 0 & 2 & 0 & 1.41 & 22.38 \\ \hline
\textbf{Case 3-2} & 338 & 573 & 1,067 & 1,647 & 1.28 & 28.26 \\ \hline
\end{tabular}
}
\begin{tablenotes}
\tiny
\item[1] NVIDIA K80 GPUs, Intel Xeon E5-2686 v4 62 G RAM
\item[2] NVIDIA GTX 1050 GPUs, Intel i7-7700HQ 32 G RAM
\end{tablenotes}
\end{threeparttable}
\label{table:per}
\end{table}

For Subcase 1 (Case 1-1, 2-1, 3-1), Table~\ref{table:per} shows that the percentage of fully connected drainage networks in Case 1-1 becomes $57.0\%$ at 50,000 epochs and that in Case 2-1 increases to 88.9\%. A smaller size of the data set by dimension reduction in Case 2-1 helps the training faster with better performance given the DCGANs architecture. 
Interestingly, Case 2-1 trained with the $\mathfrak{D}$-matrix alone provides fewer number of fully connected networks (19.5\%) than Case 1-1 (35.7\%) in the early epochs of 10,000, indicating that the connectivity structure in the original set is slowly learned with the $\mathfrak{D}$-matrix. 
On the other hand, Case 3-1 trained with the proposed connectivity-informed $\mathfrak{D}$-matrices provides a large number of fully connected drainage networks (92.6\%) even at the early epochs of 10,000, highlighting the effectiveness of our proposed approach. By providing spatial directional information into separate data channels during the training, the requirements of the fully connected network are implicitly enforced and better captured in the trained models. 

Subcase 2 (Case 1-2, 2-2, 3-2) with more complex networks generated a much smaller number of fully connected drainage networks than Subcase 1 with less complex networks.
The drainage network ($\beta = 10^{-4}$) of Subcase 2 exhibits a more complicated connection pathway along the middle flow path than the less complex network ($\beta = 10^{3}$) of Subcase 1 (Fig.~\ref{fig:network_model}), resulting in much less number of the fully connected drainage network in the generation process.
As a result, Case 1-2 trained with the network images and Case 2-2 with $\mathfrak{D}$-matrices alone generated less than 1\% of fully connected networks in all training epochs.
However, in the previous loss function analysis (Fig.\ref{fig:tr_result} (b)) both $D$ and $G$ model were trained seemingly well and the accuracy of the $D$ model increases as epochs increase.
In fact, both Cases were trained to generate the drainage network with similar shapes and patterns rather than their connectivity as shown in Fig.~\ref{fig:Gen}.
These results demonstrate that the drainage network image (Case 1-2) and the $\mathfrak{D}$-matrix (Case 2-2) alone are not enough for DCGANs to suitably reproduce the connectivity between neighboring nodes especially in the relatively complex drainage network. 
Addition information enforcing the network connectivity should be incorporated for river network generation. 

Trained with the proposed connectivity-informed $\mathfrak{D}$-matrices (Case 3), DCGANs successfully generated a number of fully connected drainage network samples with relatively large complexity. Case 3-2 shows the increase in the fully connected networks from several hundreds to 1,647 over the training. While the chance of generating fully connected network is still low (16.4\%), Figure~\ref{fig:Gen} clearly shows that a majority of fully connected networks in Case 3-2 has the connected pathway to the network outlet, while the network connectivity in Cases 1-2 and 2-2 are partly broken or short circuited with Case 2-2 being fragmented into small clusters. Practically, with the proposed approach, one may generate as many samples as possible and screen them for generating the required number of fully-connected network samples. 
These results demonstrate that the proposed connectivity-informed approach allows DCGANs to learn key physical features (e.g., connection pathway(s)) inherent in the original drainage network samples suitably and better reproduce the fully connected drainage networks than the other two cases.

\subsubsection{Evaluation of Network Similarity via Stochastic Analysis}
In this subsection, we perform a stochastic analysis of surrogate runoff response at the outlet of the network for evaluating the network complexity and similarity of the generated drainage networks. 
It is often difficult to visually distinguish one from the other in many natural and man-made drainage networks and determine their network complexity.
Alternatively, the similarity of the drainage networks can be evaluated based on the runoff at the outlet. Here we use the width function, which has been widely used as a gauge of the shape of the catchment to compare the properties of the channel network like the drainage network at various grid resolutions \cite{Rinaldo_etal_95, Fekete_etal_2001, Moussa_08}.
The width function describes the flow path from each pixel to the outlet, and consequently it depends on the geometric position of the nodes, the area drained by each node, and the distance from each node to the outlet in drainage networks \cite{Moussa_08}.
The width function can capture the essential features of the drainage network’s response so that the quality of generated networks by DCGANs is evaluated compared to the original networks simulated by the Gibbs' model. 
The width function and the area function can be differently defined based on channelization \cite{Lashermes_07}.
In this study, width function was obtained by counting the number of grid points given a distance from the outlet as 
\begin{equation}
    W(\xi) = \sum\limits_{i=1}^n s(x_{i}) 
    \label{eq:W}
\end{equation}
where, $\xi$ is the distance from the outlet along the drainage path, and $s(x_{i})$ is the number of grid points drained by each node $x_{i}$ with a distance $\xi$.

\begin{figure}
    \centering
    \includegraphics[width=0.7\linewidth]{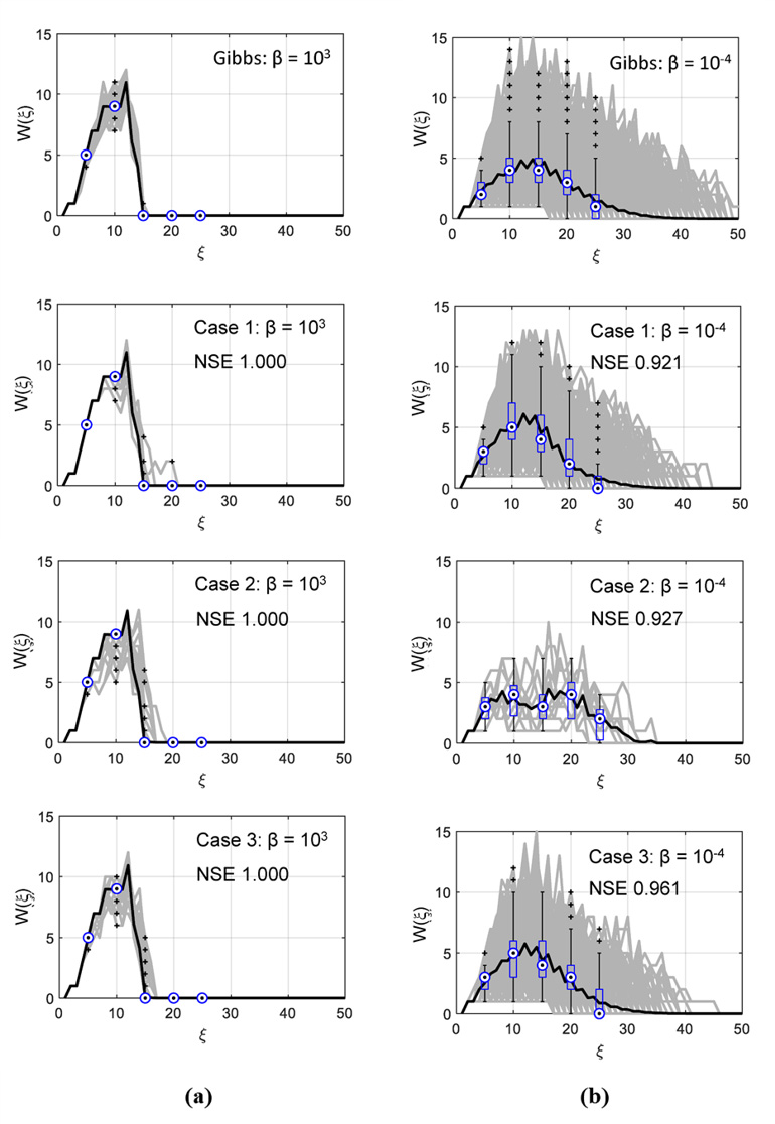}
    \caption{Comparison of width functions of the generated 1,000 fully connected networks from (a) Gibbs' model with $\beta = 10^{3}$ (Subcase 1); (b) Gibbs' model with $\beta = 10^{-4}$ (Subcase 2). The width function for Case 2-2 (Case 2: $\beta = 10^{-4}$) was obtained from only 20 generated networks because Case 2-2 hardly generated the fully connected networks.}
    \label{fig:cf}
\end{figure}

We compared the width functions obtained from Gibbs' model and 1,000 fully connected networks generated for each Case as shown in Fig.\ref{fig:cf}; the width function for Case 2-2 was obtained from only 20 generated networks because the DCGANs in Case 2-2 hardly generated the fully connected networks. 
The width functions from Gibbs’ model with $\beta = 10^{3}$ (Subcase 1) shows a higher peak and shorter flow travel stance in the runoff response. The width functions from Gibbs’ model with $\beta = 10^{-4}$ (Subcase 2) has a lower peak and long travel distance due to the complex network topology of Subcase 2. 
Although individual width functions (gray lines) of the generated fully connected networks differ slightly from the original of the Gibbs' model, the averaged width functions (solid black lines in Fig.\ref{fig:cf}) show that the Nash-Sutcliffe efficiencies (NSE) was above 0.9 in Subcase 2 and close to 1 in Subcase 1.
This means that the fully connected drainage networks generated by DCGANs have various drainage paths to the outlet with almost same complexity which has identical responses to the original networks by Gibbs' model.
In particular, Case 3 not only generated the largest number of fully connected network with various drainage paths to the outlet, but also generated networks with most similar complexity to the original drainage network.

\subsection{Efficiency}
\begin{figure}[hb]
\centering
\includegraphics[width=0.5\linewidth]{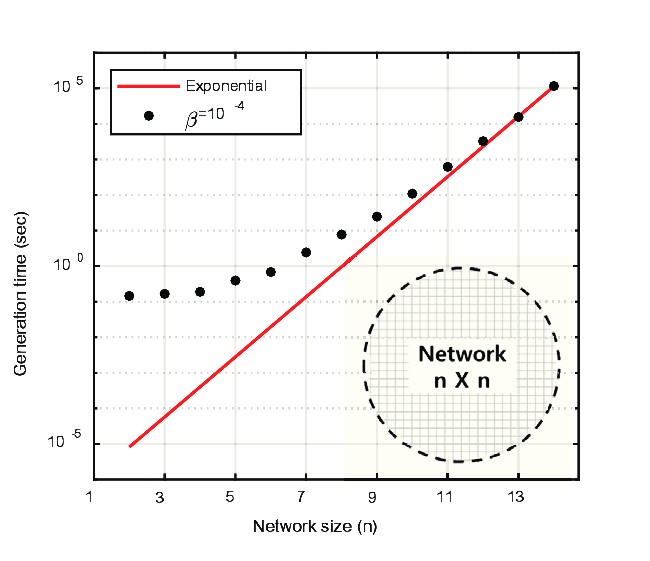}
\caption{Time spent for generating one drainage network with a size of $n$ using Gibbs’ model with $\beta = 10^{-4}$. It took more than 10 minutes to generate one drainage network with the network size, $n = 11$, used in this study.}
\label{fig:time}
\end{figure}

The stochastic network model takes a relatively long time to generate individual drainage network because of its probabilistic approach for all flow directions at each nodes of the network.
Specifically, the computation cost increases exponentially when generating more complex and bigger network.
Fig. \ref{fig:time} shows the time spent for generating one drainage network with a size of $n \times n$ using Gibbs' model with $\beta = 10^{-4}$ and the generation time of one network indeed increases exponentially with the size. 

On the other hand, DCGANs, once trained, could generate very quickly a large number of drainage networks with low computational costs.
The average time for DCGANs to generate 10,000 drainage networks were all less than one minute as shown in Table.~\ref{table:per} on a computer equipped with NVIDIA GTX 1050 GPUs and Intel i7-7700HQ 32 G RAM. 
Compared to the stochastic network Gibbs’ model, network generation time for statistical evaluation could be significantly reduced by DCGANs.
In training DCGANs, the proposed method using connectivity-informed $\mathfrak{D}$-matrices could reduce the number of training epochs required to generate fully connected networks as shown before, and increase the training speed of DCGANs due to its compact representation of network complexity and connectivity with a smaller size than the original network images (Table.~\ref{table:per}).

\section{Conclusions}
In this study, DCGANs were applied to quickly reproduce many drainage networks from the drainage network samples already generated by the stochastic network generation model, Gibb's model.
DCGANs have the promising potential for quickly generating similar network topology, as many previous studies have already shown that it could generate similar images and patterns very well.
However, DCGANs trained with the drainage network image did not properly reproduce the network connectivity inherent in the drainage network due to the complex features and patterns in the drainage network images.
The additional information of the connectivity via the drainage network sample was required for DCGANs to effectively learn and reproduce the physical information such as network connectivity and complexity in the original drainage network samples.
Hence, the DCGANs trained with original drainage network images were compared with two other cases where two different types of training samples were constructed with the directional information only ($\mathfrak{D}$-matrix) and the connectivity-informed directional information (binary matrices layers).

$\mathfrak{D}$-matrix was more effective as a training sample than the drainage network images to reproduce the connectivity in the less complex drainage network topology where many connection pathways to the central drainage path were possible. 
However, for the relatively complex drainage network samples, both DCGANs trained with $\mathfrak{D}$-matrix and network images rarely reproduce the fully connected drainage networks.
This poor performance was attributed to the fact that both drainage network images and $\mathfrak{D}$-matrix do not explicitly exhibit the spatial structural information such as the directional connectivity in the drainage network samples and DCGANs would require more training with wider and deeper neural network architectures. 
Without accounting for the connectivity between adjacent pixels (or nodes in the $\mathfrak{D}$-matrix), each pixel (node) in the network images ($\mathfrak{D}$-matrix) can take any values (directions) regardless of the value (direction) of the adjacent pixels(nodes) (i.e., high-frequency feature).
Our novel connectivity-informed method in the form of binary matrix layers performed much better than the other two cases, indicating that both directional information and their constraints on connectivity were embedded into two or three binary matrices layers (connectivity-informed $\mathfrak{D}$-matrix) so that the connectivity constraints between the directions on each node in the drainage network can be optimally stored.
In training DCGANs, the proposed connectivity-informed $\mathfrak{D}$-matrix could train DCGANs more effectively in term of accuracy and computational cost, which can be used to quickly generate many drainage networks with better representation of the network complexity and connectivity of the original drainage network sample as shown in the width function analysis in this work.

This study highlights that the generation performance of DCGANs to reproduce the structural features of images can be improved by transforming the physical information of the images (i.e., high-frequency features and connectivity between the neighboring nodes) into the efficient binary matrix layers.
Since the complex and sparse features are common in many earth and material sciences such as fractures, defects, connected pathways in porous media (e.g., pore network), and engineered features for high conductive pathways, the connectivity-informed method developed in this study can be applicable for generating these challenging multi-dimensional features in a computationally efficient way with relatively high statistical accuracy.

\section*{Acknowledgements}
Jonghyun Lee was supported by Hawai'i Experimental Program to Stimulate Competitive Research (EPSCoR) provided by the National Science Foundation Research Infrastructure Improvement (RII) Track-1: 'Ike Wai: Securing Hawai'i's Water Future Award OIA \#1557349 as well as an appointment to the Faculty Research Participation Program at the U.S. Engineer Research and Development Center, Coastal and Hydraulics Laboratory administered by the Oak Ridge Institute for Science and Education through an interagency agreement between the U.S. Department of Energy and ERDC. Yongwon Seo and Junshik Hwang were supported by Basic Science Research Program by the National Research Foundation of Korea (NRF-2016R1D1A1B03930893). This work was also supported by the Laboratory Directed Research and Development program at Sandia National Laboratories. Sandia National Laboratories is a multimission laboratory managed and operated by National Technology and Engineering Solutions of Sandia, LLC., a wholly owned subsidiary of Honeywell International, Inc., for the U.S. Department of Energy’s National Nuclear Security Administration under contract DE-NA-0003525. This paper describes objective technical results and analysis.  Any subjective views or opinions that might be expressed in the paper do not necessarily represent the views of the U.S. Department of Energy or the United States Government.

\end{document}